\documentclass[12pt]{article}
\usepackage{rotating}
\usepackage{multirow}
\usepackage{verbatim}
\usepackage{cite}
\usepackage{amsmath}
\usepackage{fullpage}
\usepackage{setspace}
\usepackage{booktabs}
\usepackage{graphicx}
\usepackage{color}
\usepackage{float}
\usepackage{nameref,hyperref}

\newcommand{\beginsupplement}{%
        \setcounter{table}{0}
        \renewcommand{\thetable}{S\arabic{table}}%
        \setcounter{figure}{0}
        \renewcommand{\thefigure}{S\arabic{figure}}%
     }

\begin {document}

\title {Embryonic Elongation controlled by Graded Cell Fate: A Molecular Dynamics Approach}

\author {Anna Mkrtchyan$^{1}$, Bertrand B{\'e}naz{\'e}raf$^{2}$, Olivier Pourqui\'e$^{3, 4}$, Paul Fran\c{c}ois $^{1}$}
\maketitle
\begin{center}
$^{1}$ Department of Physics, McGill University, Montr\'eal, QC, Canada
\\
$^{2}$  Institute of Genetics and Molecular and Cellular Biology, Strasbourg, France
\\
$^{3}$  Department of Genetics, Harvard Medical School, Boston, MA, USA
\\
$^{4}$ Department of Pathology, Brigham and Women's Hospital, Boston, MA, USA
\end{center}

\begin{abstract}
Regression of Hensen's node from anterior to posterior is driving the elongation and patterning of avian embryo body. Recent experiments link gradient of presomitic mesoderm cell motility to displacement of the node and body axis elongation. Ingression of new cells into presomitic mesoderm tissue also contributes to the process. At present, movements of presomitic mesoderm can be tracked at single cell precision. Yet, mechanisms that couple these movements to regression and axis elongation are largely unknown.
In this work we develop a computational approach to study regression of Hensen's node and the elongation of anterior-posterior body axis. Based on our simulations we propose that regression and the elongation are a result of the influx of new mesoderm cells mediated by cell density gradient. Addition of new cells leads to expansion of tissue in anterior-posterior direction (elongation) and pushes node towards posterior (regression). Motility gradient of cells further aids regression by biasing tissue expansion towards more motile posterior. We show that our model reproduces experimentally observed differences in presomitic mesoderm cell movements and regression of Hensen's node in various embryo phenotypes.
\end{abstract}

\section{Introduction}
\label{sec:intro}

Vertebrates develop with respect to anterior-posterior axis in head-to-tail manner~\cite{benazeraf2013formation}.
First, formation of the \textit{primitive streak} at posterior and its extension towards anterior breaks radial symmetry of the embryo and sets anterior-posterior axis\cite{stern2004gastrulation}.
The anterior tip of primitive streak is defined as an organizer center called \textit{Hensen}'\textit{s node} in avian embryos.
Gastrulation, the organization of three germ layers called \textit{ectoderm}, \textit{mesoderm} and \textit{endoderm}, takes place at the level of primitive streak.
Cells from dorsal layer, \textit{epiblast}, undergo epithelial-to-mesenchymal transition (EMT)~\cite{hay1995overview} and migrate ventrally through the streak to form endoderm and mesoderm; the remaining cells in epiblast form ectoderm.

\begin{figure}[h]
  \centering \setlength{\fboxsep}{0pt}%
  \setlength{\fboxrule}{0pt}%
  \includegraphics[width=\textwidth]{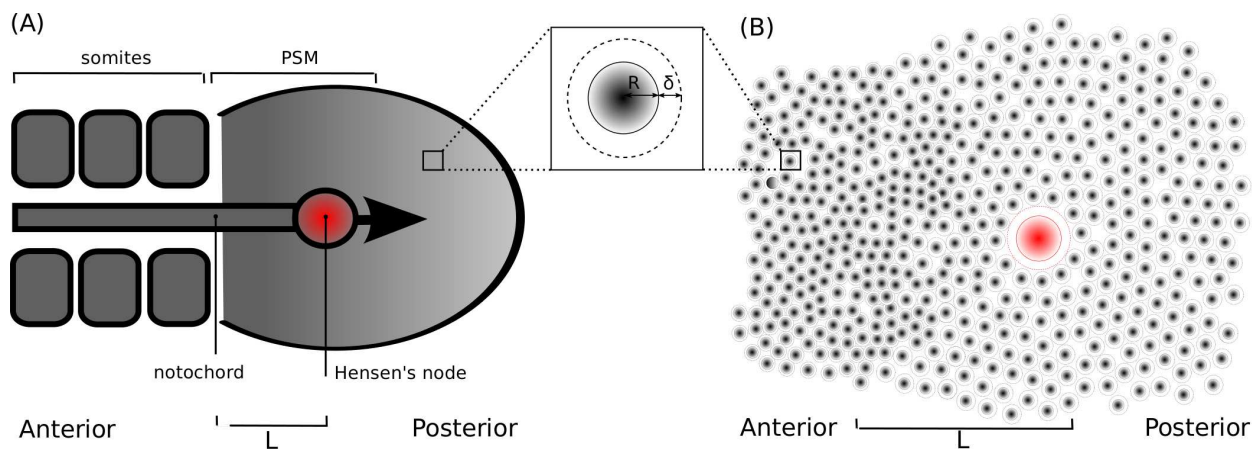}
  \caption{Modelling the elongation of avian embryo.
  (A). Schematic representation of embryo elongation accompanied by regression of Hensen's node.
  Arrow indicates direction of regression.
  Surrounding node is the presomitic mesoderm
  with cell motility and density gradient of characteristic length L.
  \textit{Inset:} Single PSM cell is approximated through an elastic soft disc. 
  Hard core radius R corresponds to cell's excluded volume.
  A shell of thickness $\delta$ defines the extent of cell's deformation due to inter-cellular interactions.
 (B). Model embryo as a collection of elastic soft discs.
 PSM tissue is modelled as a mixture of two types of cells: anterior and posterior.
 Both types have the same hard core radii R.
 Shell thickness $\delta_{A}$ and $ \delta_{P}$ for anterior and posterior cells
 are chosen such that PSM cell number density roughly follows the experimental observations~\cite{benazeraf2010random}.
}
  \label{fig:model}
\end{figure}

Gastrulation is accompanied by the elongation of anterior-posterior body axis.
As body axis elongates, Hensen's node regresses posteriorly and progressively induces the patterning of surrounding tissues~\cite{spratt1947regression}.
Cells located posteriorly to the node, in the streak and its derivatives, continue to gastrulate and ingress into embryonic mesoderm.  Once in mesoderm, they progressively differentiate into mature tissues anteriorly. For instance, ingressing cells from the anterior part of the primitive streak first form presomitic mesoderm (PSM) which later segments into somites, precursors of the future vertebraes~\cite{benazeraf2013formation}.
This process of maturation occurs anteriorly while new immature cells are continuously added at the posterior part of the tissue contributing to its elongation.
Regression of the node and sequential anterior-posterior patterning of PSM are sketched in Fig.~\ref{fig:model} (A).

A steady regression of Hensen's node implies an existence of a force directed from anterior to posterior.
B{\'e}naz{\'e}raf et al.~\cite{benazeraf2010random} performed set of experiments to identify its origins and nature.
Their works demonstrate that caudal PSM plays a key role in the posterior body axis elongation.
PSM lies laterally on both sides of embryonic midline (Fig.~\ref{fig:model} (A)).
As Hensen's node regresses towards posterior,
PSM cells left behind change their packing and motile properties and form progressively dense and less motile mesoderm~\cite{benazeraf2010random}.
The latter eventually forms somites by condensation and epithelization of mesenchymal cells (mesenchymal-to-epithelial transition)~\cite{christ1995early}. 

Cells are capable to actively change their shapes by modulating cortical forces~\cite{rauzi20114}. Collectively, such cell shape changes can result in tissue elongation through convergence-extension~\cite{keller2000mechanisms}.

The analysis of PSM cell movements relative to extracellular matrix (ECM)
showed no directional bias~\cite{benazeraf2010random}.
Individual cells perform random diffusive-like motions, much like colloidal particles~\cite{berg1993random}.
At the same time overall motion of PSM tissue is biased towards posterior. 
Significant lateral convergence that accompanies anterior-posterior elongation during convergence-extension is not observed in posterior PSM tissue elongation, which rules out convergence-extension mechanism of the elongation.

How random unbiased motions of individual cells translate to a directional motion of the entire tissue?

Collective cell movements may impose overall directionality on tissue motion~\cite{vicsek1995novel}. Cells are known to exhibit high degree of coherence at the most posterior region of the embryo, tailbud. This effect, however, is less prominent in presomitic mesoderm~\cite{lawton2013regulated}.

Another mechanism for directional tissue motion was proposed by Cai et al. in the context of neuron cell migration~\cite{cai2006modelling}.
Their modelling works of neural tissue explants suggest that presence of a gradient of motogenic signalling molecules imposes a directionality on intrinsically unbiased cellular motions.
Indeed, in the absence of motogenic substances cells spreads away from the center of the explant in a symmetric fashion.
Introducing motility promotor (inhibitor) substance to a cell population biases directionality of cell spreading towards (away) from the source of the substance~\cite{cai2006modelling, benazeraf2010random}.

Molecules of Fibroblast Growth Factor family (FGF) are known to control the motility of PSM cells in avian embryos~\cite{delfini2005control}.
Specifically, gradient of FGF8 originates around Hensen's node and establishes a decreasing gradient of PSM cell motilities in posterior-anterior direction.
Thus, FGFs can potentially act as motogenic molecules in embryos.
This gives rise to the hypothesis that directional motion of PSM tissue is an emergent property of the graded motility inflicted on PSM cells by FGFs~\cite{benazeraf2010random}.

The spread of cells in neural explants is triggered by initial inhomogeneity in cell population density imposed on the system by the nature of the experiment. 
Similarly, one can assume that stationary regression of node in embryos requires the formation and maintenance of inhomogeneous PSM density throughout the regression. Mechanisms responsible for establishment of stationary gradient of cell density leading to directional motion in embryos are not well understood.

Flux of new PSM cells is another important contributor to embryonic body development.
EMT and ingression of new cells into mesoderm is a necessary process to initiate limb budding~\cite{gros2014vertebrate}.
Suppressing the ingression of cells into PSM severely truncates the elongation of body axis in mice~\cite{boulet2012signaling}. 
It is not clear how ingressed cells contribute to the regression of the node specifically in the context of motogenic hypothesis.

In this paper we employ computational methods to address these questions above.
We start by developing a single cell based stochastic model for the elongating embryo.
Individual treatment of cells allows to manipulate their motile, packing and growth properties on a single cell level. 
It also allows to quantify the extent to which a specific mechanism influences the regression of the node.

Our simulations predict causality between the influx rate of PSM cells and the regression rate of the node.
Higher PSM influx rates correspond to higher regression rate of the node.
Interestingly, based on our results, the steepness of cell density gradient affects the influx rate of PSM cells the most.
Simulated systems with sharper density gradients have higher influx of cells added to the system.
Systems that lacked density gradient display lesser extent of node regression
even though cellular motilities were graded.
Based on our findings we propose and numerically validate the following mechanism for regression of Hensen's node.
The ability of cells to pack into more compact structures in anterior promotes the influx of the cells from posterior to anterior.
The regression of the node is then driven by excluded volume effects.
Cell motility gradient further aids node's regression by prohibiting tissue expansion towards less motile anterior and biasing it towards more motile posterior.

\section{Model and Methods}
\label{sec:model}

Experimentally observed diffusive-like motion of individual cells
has long served as a base for development of simple models,
where cells are treated as colloidal particles~\cite{anderson2007single}.
One such model was originally proposed by Drasdo et al.~\cite{drasdo1995monte},
Briefly, a cell is approximated through a spherical particle with a hard core of radius R and a shell of thickness $\delta$ as shown in Fig.~\ref{fig:model} (\textit{Inset})).
The former corresponds to the excluded volume of cell
while the latter determines the extent of cell's ability to deform.

Our model is developed based on analogous cell representation.
Essentially flat structure of the embryo allows to employ a two dimensional framework and model elongating embryo as a collection of two-dimensional elastic discs as described above performing a Brownian motion.
Since caudal PSM affects regression of the node the most, we limit our model to explicitly considering Hensen's node and surrounding PSM cells.
PSM cells are embedded into extracellular matrix (ECM). It has been shown that ECM tissue motion closely follows the motion of PSM cells~\cite{benazeraf2010random}. Based on these observations, we do not consider ECM explicitly and assume its motion is mediated by PSM cells.
The system is further simplified by disregarding any structural details of the node and considering it to be a Brownian particle of a larger size.

Below we present the set of model assumptions.

\subsection*{Cell Phenotypes.}

PSM cells in wild type embryos display different packing and motile properties depending on their position along anterior-posterior axis~\cite{benazeraf2010random}.
Moreover, set of experiments on single cells extracted from anterior and posterior sections of PSM tissue showed distinct differences in cellular behavior outside of tissue context (unpublished data by B{\'e}naz{\'e}raf and Pourqui\'e).  Based on these observations we propose a phenomenological model where PSM consists of two types of cells
with different physical properties.
The existence of distinct cell ``physical phenotypes" in mesenchyme was earlier considered in limb morphogenesis (see e.g.~\cite{damon2008limb}).
Cells in limb bud expressed higher cellular cohesivity and cell number density compared to flank mesoderm.
The gastrulation movements in Drosophila embryo requires existence of cells with different mechanical properties~\cite{rauzi2015embryo}.
We assume that anterior and posterior PSM cells are distinct physical phenotypes of PSM with different physical properties, such as size and motility. 
To account for packing differences in anterior vs posterior, posterior cells are assigned larger radii and higher diffusion coefficients compared to anterior cells (Fig.~\ref{fig:model} (B)).

In avian embryos Hensen's node acts as an organizer of embryonic development. 
We capture gradual change in PSM packing and motility by assigning a cell one of the two types based on its relative position with respect to the node according to the probability 

\begin{equation}\label{eq:prob} 
		 \phi_{A} = \left \{ \begin{array} {ll}
			0 &\mbox{if $x_{cell}\geq x_{node}$} \\
			1 & \mbox{if $x_{cell}\leq x_{node}-L$}\\
			\sim(x_{node}-x_{cell})& \mbox{if $x_{node}-L < x_{cell}<x_{node}$}
			\end{array} \right.		 
 \end{equation}

Here, $\phi_{A}$ is the probability of a cell being of an anterior type,
$x_{cell}$ and $x_{node}$ are positions of the cell and the node along the anterior-posterior axis,
and L is the characteristic length of the density gradient (see Fig.~\ref{fig:model} (B)).

\subsection*{Cell-cell interactions.}

Motions of cells inside tissue are influenced by their interactions with neighbouring cells.
Cells in vicinity may form adhesive bonds to maintain the integrity of the tissue.
They exert elastic repulsive forces on neighbour cells as a response to compression.
Drasdo et al.~\cite{drasdo2007role} explored several variants of interaction potentials between cells
and found that the exact form of potentials does not affect qualitative description of developmental processes.
For simplicity, we model cell-cell repulsion and adhesion with linear spring-like forces.
Caudal PSM cells are only loosely connected within tissue.
Anterior PSM cells, on the other hand,
change their adhesive and migratory properties and eventually condense into somites~\cite{gossler1997somitogenesis}.
With the above,
we assume that a pair of anterior and posterior 
or two posterior cells interact solely through repulsive elastic forces
modelled by linear springs.
To facilitate the tendency of anterior cells to condense,
in addition to the elastic repulsion, we also consider weak cell-cell adhesion force between two anterior cells in vicinity.
Detailed description of inter-cellular forces and the choice of the interaction parameters is presented in Sec. \nameref{sec:forces} and ~\nameref{sec:params} correspondingly.

\subsection*{Addition of New Cells.}

Influx of new PSM cells is maintained by ingression of cells from epiblast as well as by cell proliferation in PSM.
Understanding origins and nature of newly added PSM cells is beyond the scope of this paper.
Here, we are only concerned about the fact that there is an ongoing addition of mass (new cells) to the system.
As such, we do not distinguish between processes of ingression and cell proliferation and model the
addition of new cells to the system via growth and division of the existing ones.
Cells grow by means of small stochastic increments of their radii.
Once cell's area is doubled, it is divided through randomly oriented division line.
During the division, cell is replaced by two daughter cells
of same type and physical properties as their parent.

Caudal PSM has the most prominent effect on regression of the Hensen's node.
For computational efficiency,
we neglect the proliferation outside of caudal PSM
assuming it has only a small influence on regression.
The region where new cells are added is defined relative to the instantaneous positions of the node. 
In anterior, cells within a distance $GR_{A}$=L (same as characteristic distance of density gradient)
from Hensen's node grow,
provided their local number density does not exceed a threshold value $\rho_{thres}$.
The latter requirement makes sure that the growth of the cell is suppressed in the crowded environments, reducing the unrealistic overlap of cells in the system.

Similarly, in the posterior, 
cells that are within $GR_{P}$ distance from the node are candidates for the growth.
Adding new cells in posterior is necessary to maintain the existence of tissue posterior to the node throughout simulations. From a biological point of view, this posterior growth regions may be considered as the portion of the primitive streak that contributes to the influx of PSM cells into the system.

\subsection*{System Dynamics.}

Langevin equation of motion for the \textit{i}-th cell takes form

 \begin{equation}\label{eq:motion}
		 \begin{array}{l}
		  	m_i\cdot\frac{\vec{r_i}}{dt}=\vec{F_{i}}/\gamma_i+\sqrt{D_i}\cdot\vec{\eta_i}
    	 \end{array}
   \end{equation}
   
Here, $m_{i}$ and $\vec{r_i}$ are the mass and the position of the cell,
$\vec{F_{i}}=\sum\limits_{j}\vec{F}_{ij}^{HC}+\sum\limits_{j}\vec{F}_{ij}^{SC}+\sum\limits_{j}\vec{F}_{ij}^{Attr}$ is the total inter-cellular force acting on a given cell,
and $\eta_i$ is the Gaussian noise.
Similar to thermal systems, damping and diffusion coefficients  $\gamma_i$  and $D_{i}$ in cellular systems relate as $D=F_{T}/\gamma$, where $F_{T}$ is the analogue of the thermal energy $K_{BT}$~\cite{drasdo1995monte}.
We consider cell type dependent diffusion coefficients to reflect differences in anterior and posterior cell motilities.

\subsection*{Simulation Procedure.}

We follow time evolution of the system by solving the system of equations of motions (Eq.~\ref{eq:motion}) with the molecular dynamics simulation methods.
Cells are placed in simulation box with elastic boundaries mimicking lateral and anterior-posterior boundaries of embryo. If cell moves outside of the boundary box, a linear spring-like force directed towards the simulation box acts on it. 
We define lateral boundaries to be stiffer than the anterior/posterior boundaries.
The choice of boundaries is motivated by
a relatively unrestricted extension of the embryo in the anterior-posterior direction.
The existence of lateral plates, on the other hand, prevents expansion of embryo in the lateral direction.
The choice of the position and the strength of lateral boundaries is arbitrary. We find that our system does not display strong dependence on the boundary conditions (See \nameref{S3_Fig}).

Our simulations start from 500 cells
that are assigned types according to the probability as defined in Eq.(~\ref{eq:prob}).
At each time step we identify cells that are candidates for growth as described in previous section. Radii of growing cells are increased at random
which introduces asynchrony in cell cycles.
Every $T_{div}$ steps we check cells' areas and the ones with the doubled area are replaced by two cells of the same type as their parent.
Such division algorithm may create an overlap of daughter cells, however the system is equilibrated within few time steps.

Cell types depend on their relative positions from the node.
At fixed time steps $T_{type}$ we estimate the position of the node on anterior-posterior axis 
and re-evaluate the probability distribution of cell types.
Cells are then assigned new types according to the updated probability.

Following this procedure, we grow systems of up to few thousand cells (See \nameref{S1_Movie},  \nameref{S2_Movie} and  \nameref{S3_Movie})

Due to the spherical symmetry of model cells anterior-most and posterior-most regions show a high degree of order, which is not observed in experiments. This is an artefact of the model since cells in real tissues constantly change their shapes as a response to changing environment which we do not consider here. 
To ensure this artificial ordering does not have prominent effect on regression of the node,
we introduces a disorder in the packing of cells in by varying hard core radii of the cells.
Indeed, variability of cells' sizes did result in the less ordered packing of cells in system,
and the qualitative behaviour of the model system remained the same (See \nameref{S2_Fig} and \nameref{S7_Movie},  \nameref{S8_Movie}, \nameref{S9_Movie}).

\subsection*{Parametrization.}

We discuss detailed parametrization and the choice of numerical values for model and simulation parameters in Sec. \nameref{sec:params}.
Briefly, our model is dimensionless where cell's mass, hard core radius and metabolic energy are set to unity of mass, length and energy accordingly. All results  are reported in reduced units. 
Corresponding real values can be estimated once the appropriate conversion base between reduced and real units is established.
We choose real values for the mass of eukaryotic cell (m=10\textsuperscript{-12} kg), cell's metabolic energy (E=10\textsuperscript{-16} J~\cite{drasdo2007role}) and the characteristic length of PSM density gradient (L=500$\mu$m~\cite{benazeraf2010random}) as a conversion base.
Then, time unit is evaluated as $[t]=\sqrt{[m][l]^2/[E]}$. This leads to following real values for the base conversion units: [m]=10\textsuperscript{-12} kg, [l]=20$\mu$m and [t]=10\textsuperscript{-3}sec. 
To get a better insight into the quantitative differences between the model and real system we compare estimated real values of model parameters with the appropriate experimentally measurable quantities.
Comparison reveals that model cells are roughly two times larger than the real cells and that for computational efficiently the dynamics of the system is accelerated by a factor of 10\textsuperscript{6}. Such scaling allows us to keep the size of simulated system and the simulation times relatively small. 

\section{Results}
\label{sec:results}

\subsection*{Trajectories, Motility and Packing of Model Cells.}

We start by examining cell motions and their packing densities for three biologically relevant PSM phenotypes expressing different FGF concentration profiles.
Wild type embryos are characterized by a gradient of FGF with the source located at Hensen's node.
FGF gain-of-function (loss-of-function) phenotypes over-express (block) FGF signalling in PSM cells.
We consider FGF implicitly through appropriately chosen distribution of cell types in the model.
Wild type embryos are modelled by considering a system
that consists of a mixture of anterior and posterior cells.
Model gain-of-function phenotypes consist of entirely posterior, more motile and loosely packed cells, whereas loss-of-function phenotypes consist of less motile and densely packed anterior cells.
Simulated elongation of wild type, FGF gain-of-function and FGF loss-of-function phenotype embryos are shown in \nameref{S1_Movie}, \nameref{S2_Movie} and \nameref{S3_Movie} correspondingly.

\begin{figure}[h]
 \centering \setlength{\fboxsep}{0pt}%
  \setlength{\fboxrule}{0pt}%
  \includegraphics[width=\textwidth]{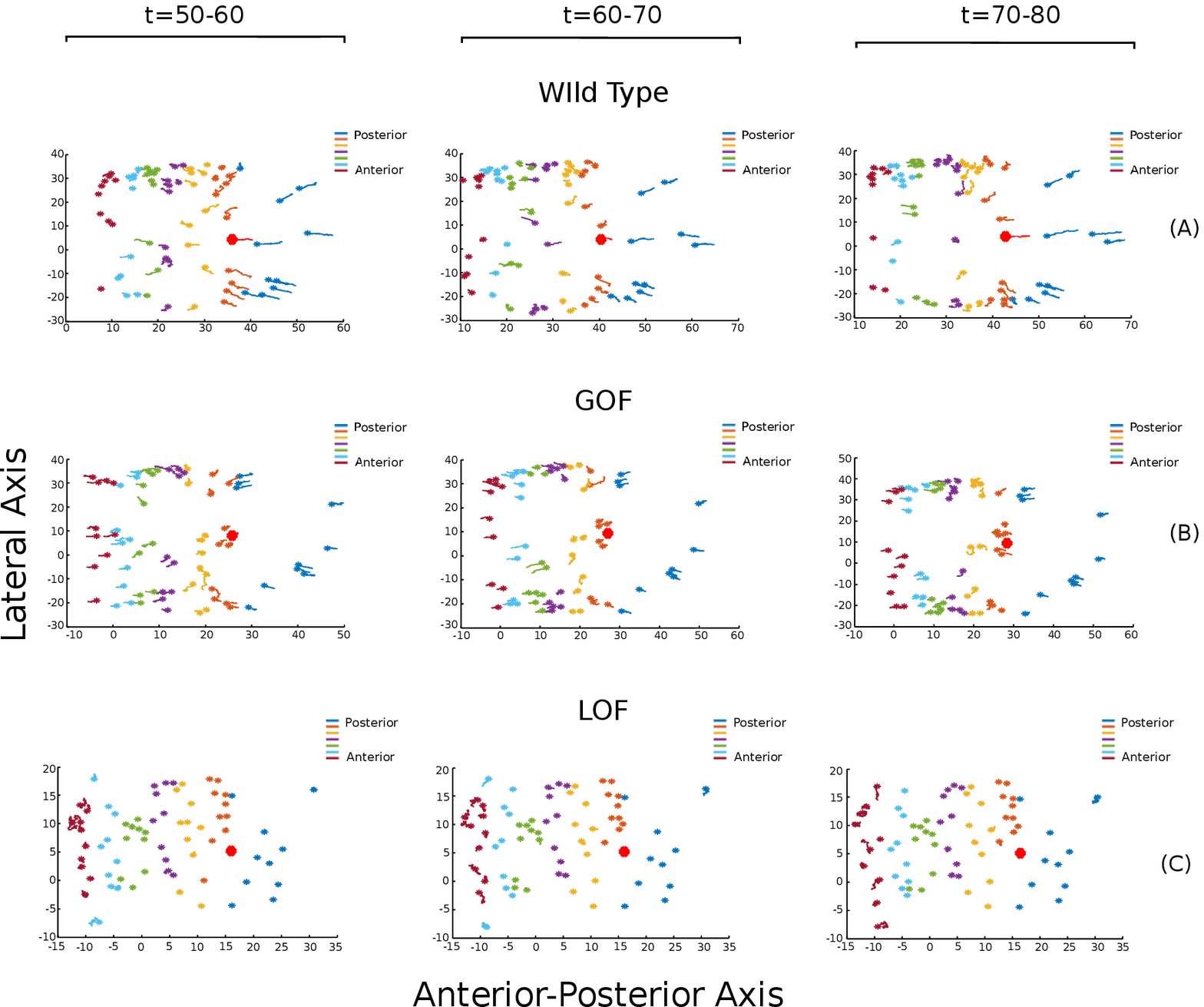}
 \caption{Trajectories of selected cells for three simulated phenotypes.
  Presented are (A) wild type, (B) FGF gain-of-function and (C) FGF loss-of-function model systems.
  Tissue is subdivided into sections along anterior-posterior axes. Different sections are color coded.
  Cells are selected at random and their initial positions are marked with asterisk.
 The trajectory of Hensen's node (red) is also shown for the reference.
 Cell trajectories are estimated over different time intervals.
}
  \label{fig:traj}
\end{figure}

First, we compare model cell movements with experimentally observed PSM deformation~\cite{benazeraf2010random}.
Simulated trajectories of few cells picked at random are shown in Fig.~\ref{fig:traj}.

We notice that our model captures the key features of cell trajectories in different phenotypes.
Simulated wild type embryos (Fig.~\ref{fig:traj} (A)) exhibit distinct regions where cell motions have different directionality.
Cells around and immediately anterior to the node move predominately towards posterior.
More anteriorly, we observed regions where cells do not seem to move in any apparent direction and regions where cellular motions are biased towards anterior.
Our results are analogous to experimentally observed trajectories of cells from various sections of PSM in wild type embryos~\cite{benazeraf2010random}.
Trajectories acquired from different simulation time intervals confirm
that this behaviour persists throughout the entire simulation.

Consistently with experiments, simulations of FGF gain-of-function system (Fig.~\ref{fig:traj} (B)) show little difference between motility of the cells in the anterior and posterior regions of tissue.
Moreover, the extent of the posterior cells' displacements is noticeably smaller.
Finally, we do not observe any significant displacements of cells in simulated FGF loss-of-function case (Fig.~\ref{fig:traj} (C)), similar to experiments.

\begin{figure}[h]
 \centering \setlength{\fboxsep}{0pt}%
  \setlength{\fboxrule}{0pt}%
  \includegraphics[width=\textwidth]{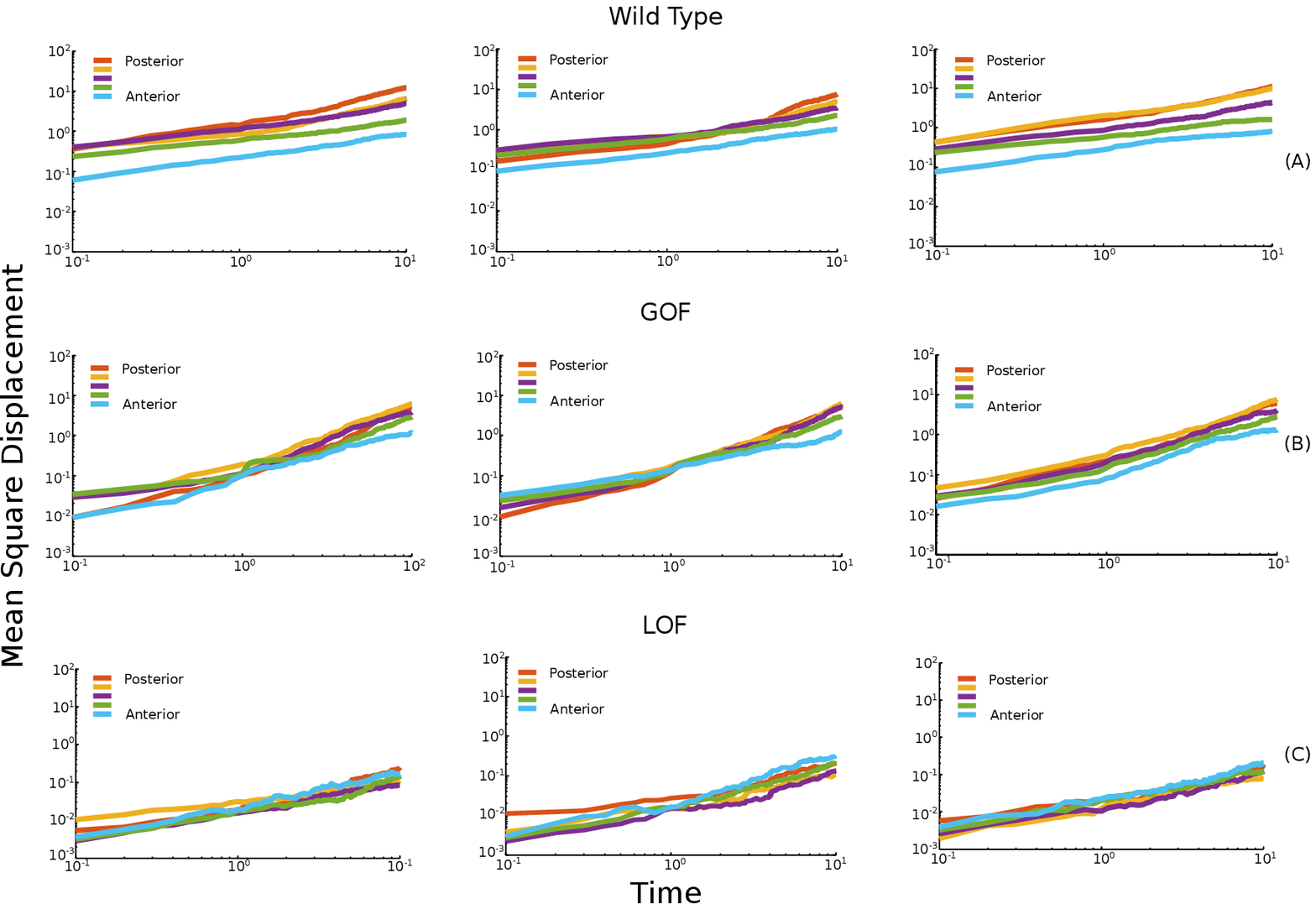}
  \caption{ Mean square displacements of cells for three simulated phenotypes (log-log plot).
  Mean square displacements are estimated for  (A) wild type, (B) FGF gain-of-function and (C) FGF loss-of-function model embryos.
  Data taken from the different sections of the tissue along anterior-posterior axis are color coded.
  Total displacement of a given section is subtracted from the displacements of individual cells prior to the estimation of cells mean square displacements.
  Time intervals over which MSD are estimated are taken small enough
  to ensure cells stay within the same section.
  While there are some variations in the MSD for all three phenotypes,
  wild type model system exhibits steeper decrease of cell motility in posterior-to-anterior direction.
  }
  \label{fig:msd}
\end{figure}

Next, we quantify the anterior-posterior profile for cell motilities inside the tissue.
Analogous to the experimental procedure~\cite{benazeraf2010random},
we first estimate cell motions relative to the tissue deformation
by subtracting total displacement of tissue
from the displacement of individual cells.
Afterwards, the mean squared displacements (MSD) of cells are estimated.
Time intervals over which MSD is calculated are taken small enough to ensure cells mostly stay within the same region relative to the node.
The results are averaged over all cells in the given section for the given time interval and are shown in Fig.~\ref{fig:msd}.

All three cases demonstrate variability in MSD,
with more anterior regions having lesser degree of cellular motility.
However, wild type embryo displays steeper gradient of motility compared to the other two cases
as observed in experiments.

\begin{figure}[h]
 \centering \setlength{\fboxsep}{0pt}%
  \setlength{\fboxrule}{0pt}%
  \includegraphics[width=\textwidth]{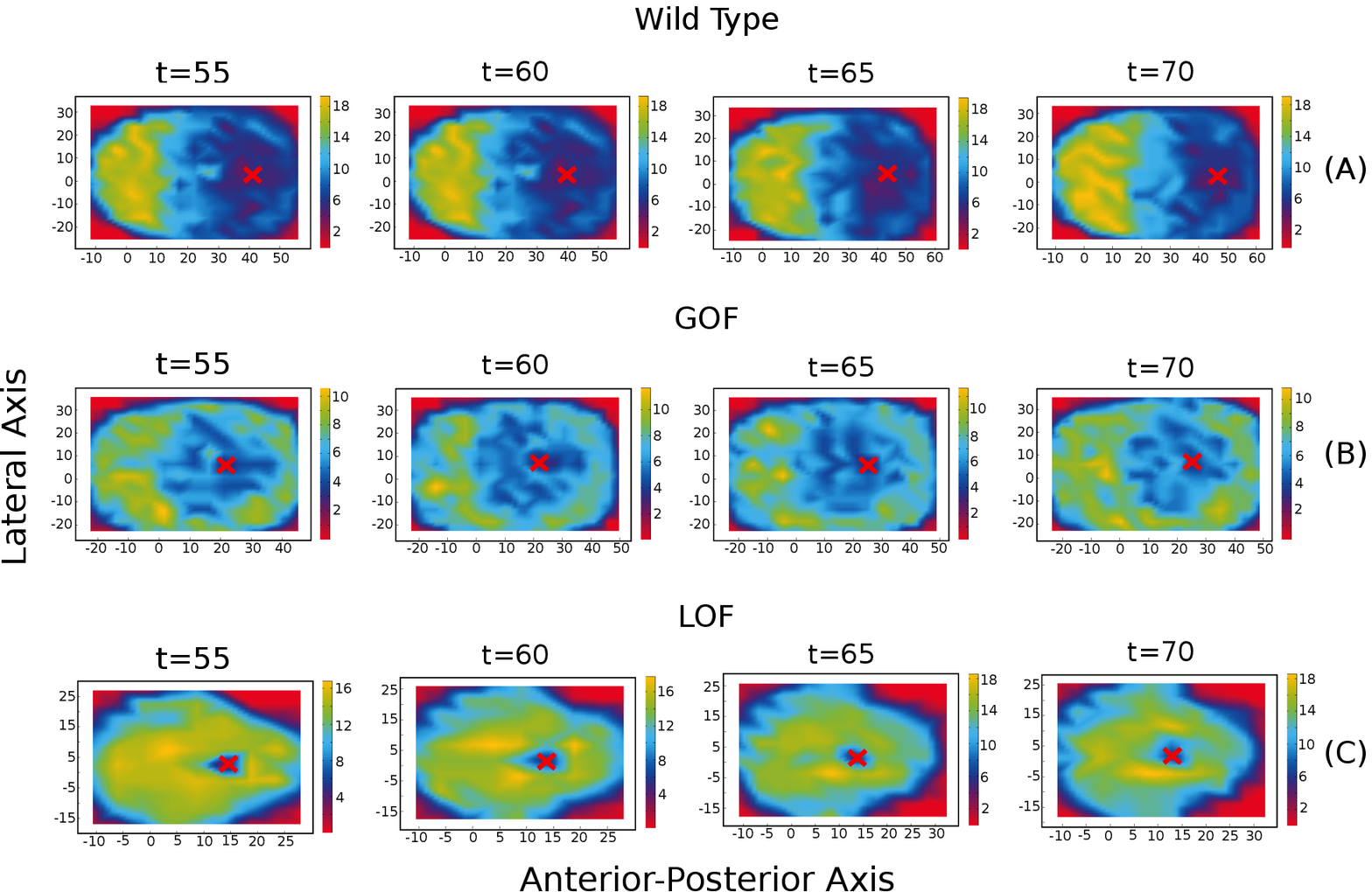}
  \caption{ Cellular number density of three simulated phenotypes.
  Cell number densities for (A) wild type, (B) FGF gain-of-function and (C) FGF loss-of-function model systems are estimated at different time points.
  Simulated wild type tissue qualitatively reproduces the gradient of cell packing density
  in anterior-posterior direction.
  FGF gain-of-function phenotype and FGF loss-of-function phenotypes
  both display relatively uniform cells packing.
  The approximate positions of the Hensen's node are marked by red.
  }
  \label{fig:dens}
\end{figure}

Another experimentally observed difference between three tissues expressing different levels of FGF is the packing densities of cells in anterior vs posterior.
To estimate cell number densities,
simulated tissue is divided  into square grid
and number of model cells in each grid cell is calculated. 
The results are then smoothed out by linear interpolation.
As one can see from Fig.~\ref{fig:dens},

Consistent with experiments, our model reproduces the gradient of cellular number density for wild type embryos.
FGF gain-of-function as well as FGF loss-of-function phenotypes feature relatively uniform packing,
with overall higher packing densities in loss-of-function case, compared to gain-of-function case.

Finally, evaluating density and motility gradient profiles at different time intervals reveals that they remain relatively stable throughout simulations (See Figs.~\ref{fig:msd},~\ref{fig:dens}). Thus, our model provides a mechanism to maintain inhomogeneous density of PSM cells throughout embryo elongation (See \nameref{S4_Movie}, \nameref{S5_Movie} and \nameref{S6_Movie}).

\subsection*{The Graded Motility and Density of PSM Cells Along With the Influx of PSM Cells is Sufficient to Account For the Regression of Hensen's Node.}

As we have demonstrated in the previous section, our two-type cell model is sufficient to
qualitatively reproduce characteristic density and motility profiles as well as the individual cell motions of avian PSM tissues.

\begin{figure}[h]
  \centering \setlength{\fboxsep}{0pt}%
  \setlength{\fboxrule}{0pt}%
  \includegraphics[width=\textwidth]{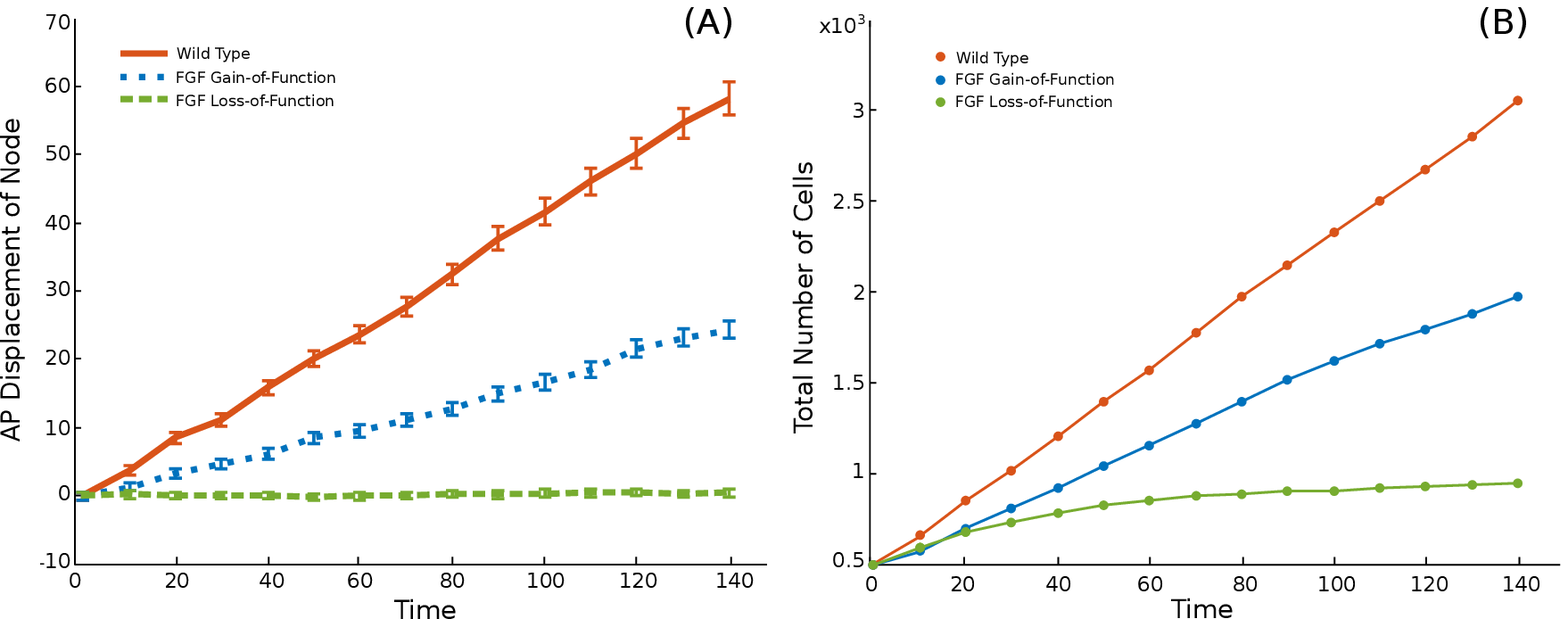}
  \caption{ Causality between cells added to system and anterior-posterior displacement of Hensen's node. (A) Anterio-posterior displacement of the node.
 Wild type embryo exhibits the highest motility of the node.
 Simulations reproduce decrease in the regression of the node for both FGF gain-of-function
 and FGF loss-of-function cases. Error bars correspond to the standard deviation of node's displacement over 1000 simulation time steps.
 (B) The extent of Hensen's node displacement correlates with the increase of the total number of cells in simulated system.
 The rates are which cells are added to the system are higher in the wild type embryos,
 followed by FGF gain-of-function and FGF loss-of-function phenotypes.
  }
  \label{fig:ap_node}
\end{figure}

In this section we quantify displacements of Hensen's node for the three phenotype tissues.
Experimental observations show that extent of node's regression is most notable in the wild type embryos.
Both FGF gain-of-function and loss-of-function phenotypes have significantly slower regression rates.
We tracked the displacement of the node by recording its  position along the anterior-posterior axis at fixed time steps for three simulated phenotypes. 
The results are shown in Fig.~\ref{fig:ap_node}.

Similar to experiments, wild type embryos demonstrate the fastest regression rates, whereas
both FGF gain-of-function and FGF loss-of-function have lesser degree of regression of Hensen's node.
We estimated the regression rate of the model node in the wild type embryos to be about $30\mu m/hr$,
which differs from the experimentally observed regression rates of the node by factor of three.
Provided cell physical properties in our model are defined up to an order of magnitude, we find that our model matches the experimental observations of the extent of node's regression quite well. 
Moreover, as we show in \nameref{S4_Fig}, node's regression rate can be easily tuned  by changing the influx rates of cells. 

Our findings reinforce the earlier proposed importance of the gradient profiles and the influx of PSM cells and suggests that these mechanisms alone might be sufficient to explain the observed regression of the node.

\subsection*{The Extent of the Hensen's Node Regression Correlates With the Steepness of the Cell Density Gradient.}

\begin{figure}[h]
 \centering \setlength{\fboxsep}{0pt}%
  \setlength{\fboxrule}{0pt}%
  \includegraphics[width=\textwidth]{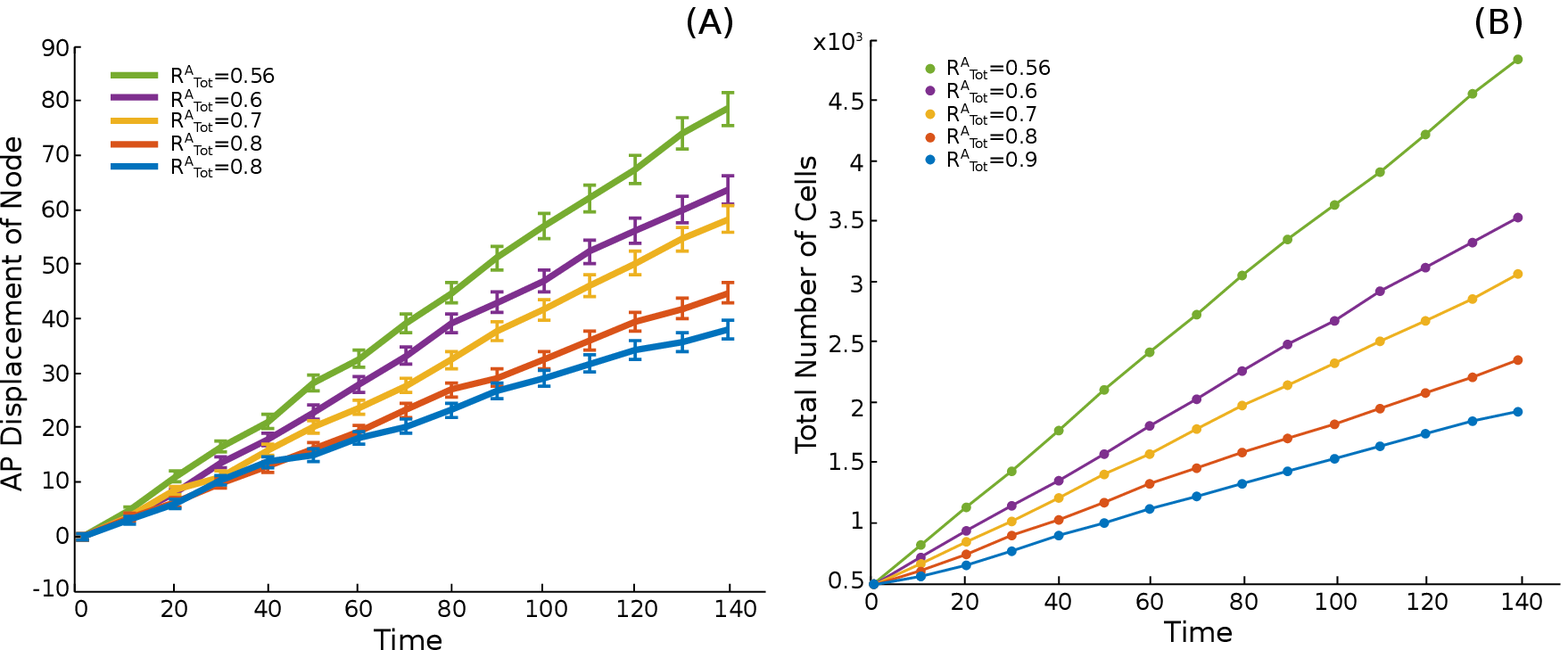}
  \caption{ Relation between the difference in anterior and posterior cells sizes,
  the anterior-posterior displacement of Hensen's node and the total number of cells in the system. 
    (A) The increase in the differences between anterior and posterior cells sizes correlates with the increase of the displacement of Hensen's node. Error bars correspond to the standard deviation of node's displacement over 1000 simulation time steps.
     (B) The increase in the differences between anterior and posterior cells sizes correlates with the increase of the total number of cells in the system.
     Data for different ratios between anterior and posterior cell radii are colour coded as follows: 
   $R_{Tot}^{A}$=0.56, $R_{Tot}^{P}$=0.9 (green),
   $R_{Tot}^{A}$=0.6, $R_{Tot}^{P}$=0.9 (purple),
    $R_{Tot}^{A}$=0.7, $R_{Tot}^{P}$=0.9 (yellow),
    $R_{Tot}^{A}=0.8$, $R_{Tot}^{P}$=0.9 (red),
    $R_{Tot}^{A}$=0.9, $R_{Tot}^{P}$=0.9 (blue).
  }
  \label{fig:vary_size}
\end{figure}

Interestingly, there is a causality between the addition of new cells and the extent of the Hensen's node displacement (Fig.~\ref{fig:ap_node}).
Both FGF gain- and loss-of-function model phenotypes display reduced growth rates compared to model wild type embryos.
The difference between model wild type embryos and the FGF phenotypes lies in the existence of gradients in cell packing density and motility,
which suggests that graded profile of the density and/or motility promotes growth (influx) of cells in system.
To confirm out assumption we gradually decrease the difference between anterior and posterior
cell radii and observe the changes in the displacement of the node and the total number of cells in the system (Fig.~\ref{fig:vary_size}).

As seen in Fig.~\ref{fig:vary_size} the greater difference between the anterior and posterior cell sizes promotes the growth rate of the system
and, consequently, the faster anterior-posterior displacement of  Hensen's node.

\begin{figure}[h]
  \centering \setlength{\fboxsep}{0pt}%
  \setlength{\fboxrule}{0pt}%
  \includegraphics[width=0.7\textwidth]{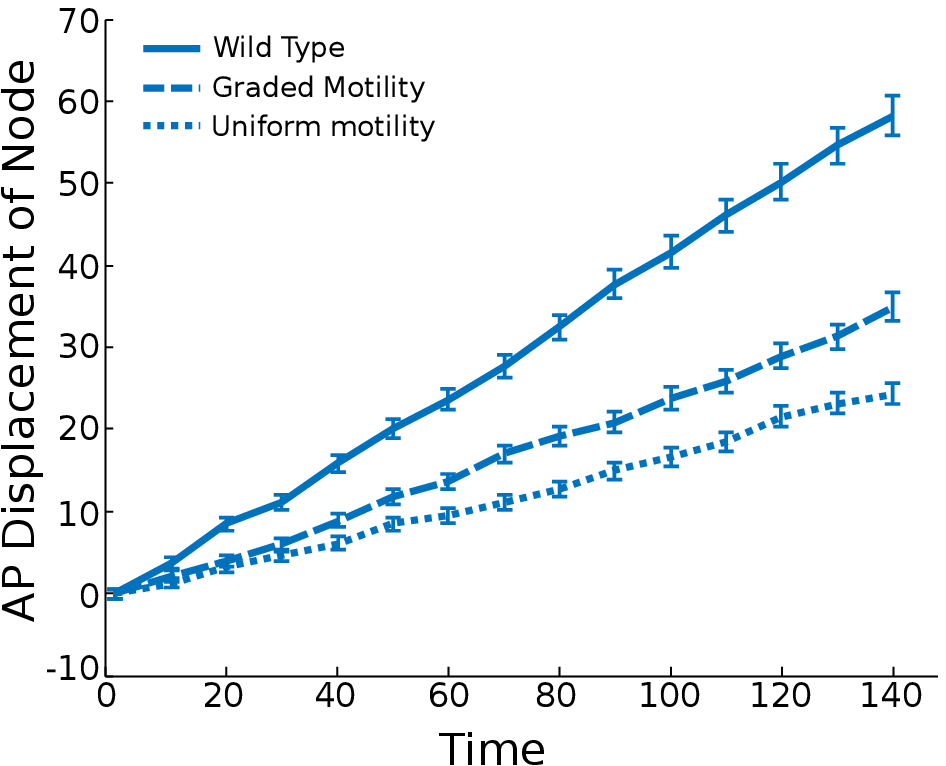}
  \caption{ The effect of the explicit diffusion gradient on the displacement of Hensen's node.
  The anterior-posterior displacements of the node in the system that has same cell sizes and explicitly introduced motility gradient analogous to wild type embryo (referred as graded motility) is compared to the FGF gain-of-function (referred as uniform motility) and wild type embryos. 
Introducing explicit gradient of motility shifts the displacement of the node towards posterior.
This shift is more prominent when density gradient is introduced as well.
  }
  \label{fig:grad_effect}
\end{figure}

To investigate the influence of explicit diffusion gradient on Hensen's node displacement,  
we considered a system that consists of posterior cells that are assigned diffusion coefficients $D_A$ and $D_P$ based on the probability~(\ref{eq:prob}).
Such system lacks density gradient observed in wild type embryos but has an explicitly introduced gradient of cell motilities as in wild type embryos. 
We then compared the node's regression in the above-mentioned system with 
simulations of wild type embryos (both density
and motility gradients are explicitly considered) and FGF gain-of-function (density gradient is absent and all cells are assigned same motility).
As we can see from Fig.~\ref{fig:grad_effect}, introducing cell motility gradient to otherwise uniform system contributes positively to regression of the node.
However, the increase in the node's regression rate is relatively small.
Introducing the density gradient along with motility gradient significantly increases regression rate.

\section{Discussion}
\label{sec:conclude}

Recent experimental works identify presomitic mesoderm as a key player in the elongation of anterior-posterior body axis and regression of Hensen's node in avian embryos.
Tracking of individual cell motions couples gradient of cell motilities to the rate of regression\cite{benazeraf2010random}. 

\begin{figure}[h]
  \centering \setlength{\fboxsep}{0pt}%
  \setlength{\fboxrule}{0pt}%
  \includegraphics[width=\textwidth]{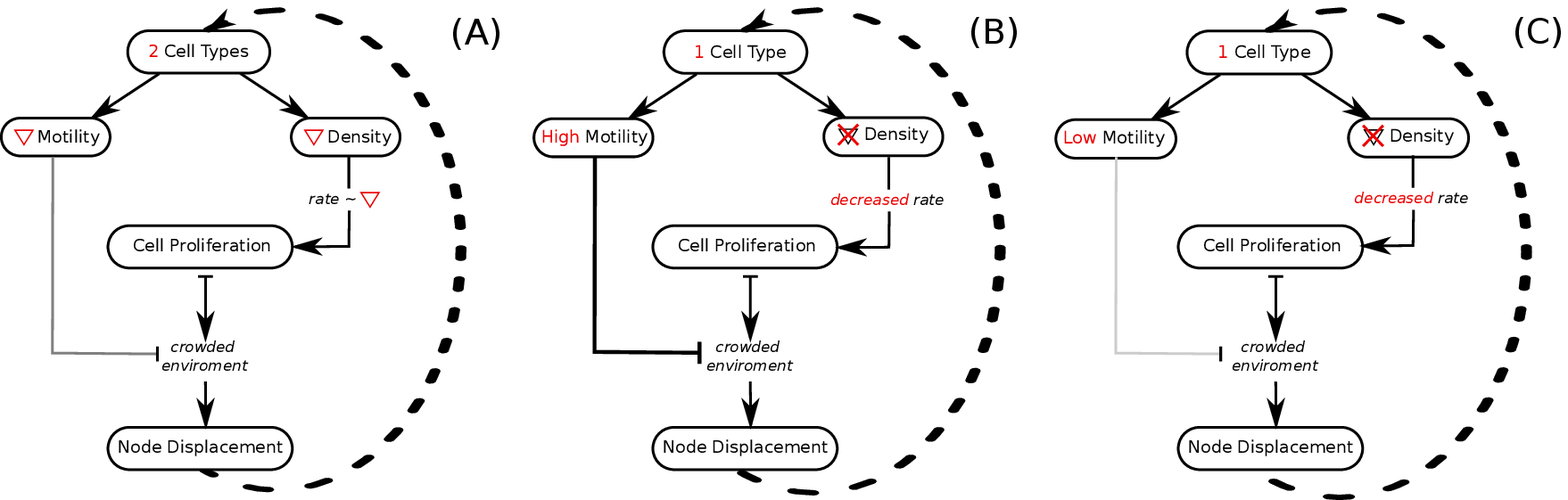}
  \caption{Proposed mechanism for stationary regression of Hensen's node.
  Solid arrows specify positive feedback. Blunt arrows specify negative feedback with the higher strength of the feedback indicated by darker color. Dashed arrow specifies re-establishment of cell types based on their updated positions relative to the node.
  (A) In wild type embryos existence of two cell types establishes cell density gradient which has a positive contribution to addition of new cells.
  Addition of new cells increases cell crowding in PSM tissue anterior to node which leads to its expansion. Motility of cells counteracts cell crowding effects and creates favourable conditions for tissue expansion. Due to gradient of cells motilities in wild type embryos the expansion of tissue and regression of node are biased towards more motile posterior.
(B) In the FGF gain-of-function embryos the absence of density gradient reduces cell addition rates. Higher cell motilities everywhere aid to tissue expansion in both anterior and posterior directions and eliminate bias in cell movements towards posterior observed in wild type embryos. These two factors lead to a reduction in node's regression rate.
(C) FGF loss-of-function embryos, in addition to the absence of gradients of cell motility and density,
have also a reduced motility of all cells.
In contrast to FGF gain-of-function embryos,
where high cell diffusion rates of cells counteract the crowding effects,
cells with low motilities stay clustered. High local cell densities penalize cell proliferation rates,
which further decreases the regression rate of the node. 
}
  \label{fig:scheme}
\end{figure}

In the attempt to understand how cells movements induce regression we have developed a single cell based model that accounts for graded density and motility of PSM tissue and incorporates influx of new cells.  Based on our studies, we proposed regression mechanism which is summarized in Fig.~\ref{fig:scheme}.
 
In wild type embryos (Fig.~\ref{fig:scheme} (A)) the existence of two types of cells with different packing and motile properties establishes gradients of motility and density (Figs. ~\ref{fig:msd} (A) and ~\ref{fig:dens} (A)).
As demonstrated by simulations,
gradient of cell packing density influences cell proliferation.
Embryos featuring steeper density gradients have higher rates of cell addition(Fig.~\ref{fig:vary_size}).
From a physical point of view, the ability of cells to pack into denser configurations in anterior can effectively decrease cell density in posterior and create space for addition of new cells via ingression or cell proliferation.
As new cells are added to the system, regions anterior to the node become increasingly more crowded which eventually leads to the expansion of the tissue through excluded volume effects.

Cell motility counteracts crowding effects and creates favourable conditions for tissue expansion.
In wild type embryos motility of cells decreases from posterior to anterior biasing 
motion of PSM tissue along with Hensen's node towards more motile posterior, by effectively prohibiting the expansion of the tissue in anterior direction (Fig.~\ref{fig:grad_effect}).
After assuming its new position along anterior-posterior axis,
Hensen's node re-establishes the gradients of cell motility and density (Fig.~\ref{fig:ap_node} (A)).

Both gain-of-function (Fig.~\ref{fig:scheme} (B)) and loss-of-function (Fig.~\ref{fig:scheme} (C)) model phenotypes
flatten density and motility gradients. The absence of cell density gradients decreases the rate of cell addition reducing regression and elongation rates in both phenotypes. 

Further difference between regression rates of FGF gain-of-function and FGF loss-of-function phenotypes lies in the difference of their cell motilities. High cells motilities everywhere in FGF gain-of-function phenotypes counteract cell crowding effects and promote tissue expansion in both anterior and posterior directions. This eliminates the bias of PSM tissue motion towards posterior due to motility gradient and further reduces regression rate of the node. 
Due to low cell motilities in FGF loss-of-function embryos cells form crowded environment early in simulations which severely reduces addition of new cells and, hence, regression rates of the elongation and regression.

Within the scope of our hypothesis we were able to reproduce regression rates of Hensen's node in wild type, FGF gain-of-function and FGF loss-of-function phenotypes up to an order of magnitude. Regression rates in the model depend on physical properties of PSM tissue, such as difference between cell sizes in anterior and posterior, influx rates of new cells into PSM, etc. Due to the lack of precise experimental values, numerical values of model parameters are defined up to an order of magnitude. Future experimental measurements of relevant PSM properties as well as careful estimates of PSM cell proliferation and ingression will aid to more accurate estimates of node's regression rate. 

While a flat structure of chick embryo allows to develop a two-dimensional model, more accurate quantitate description of the system would require to extend it to three-dimensional case.  The key physical drivers described here  (gradient of density and motility) would still remain in three dimensions, so that the main interest of 3D simulations would be to study   elongation of the embryo in the context of multiple interacting tissues. The extension is in principle straightforward by including  an approximation of a single cell through an elastic spherical particle. While such model would provide better insight into regression, it would also be much more computationally expensive, and two dimensional simulations are better suited to identify first principles.

We have used here a molecular dynamics approach to describe cell motion and  tissue elongation. Such cellular models allow for direct comparison to experimentally measured features (such as cell trajectories, mean square deviations, cellular density), corresponding to a ``microscopic'' scale .  Control of the microscopic features by a morphogen gradient then gives rise to a complex ``macroscopic'' emergent behaviour. The interplay of the microscopic/macroscopic scales is not trivial since presumably many different microscopic mechanisms could give rise to some macroscopic elongation. Simulations of mutants and of microscopic features allow for a better biological understanding of the origin of embryonic physics. Insights from this approach can later give rise to more continuous and biologically realistic description, that would be necessary to scale-up simulations of embryogenesis.

\clearpage
\bibliographystyle{unsrt}
\bibliography{references}

\clearpage
\beginsupplement

\section*{Supplementary Text}
\subsection*{Inter-Cellular Forces}
\label{sec:forces}

In the main text we introduced two types of cells,
anterior and posterior.
Cells in posterior are loosely connected within PSM tissue.
As cells are displaced towards more anterior positions, 
they start expressing adhesive properties and condense into somites.

For simplicity we assume that all cells interact through harmonic-like potentials.

(a) \textbf{\textit{Posterior}-\textit{posterior} cell interactions}: 
Posterior cells are approximated as elastic discs.
At large separations, the interaction between two posterior cells is negligible.
As the distance between two cells decreases, they come into contact and compress
which results in an elastic repulsive interaction. 

Let $R_i$, $R_j$ be the hard core radii of $i$th and $j$th cells and $\delta_i$, $\delta_j$ be the ranges over which they can be compressed.
The interaction force between two posterior cells is defined as
 \begin{equation}
		F_{ij} = \left \{ \begin{array} {ll}
			-\frac{K_{HC}*R_{HC}-K_{SC}*R_{SC}}{R_{HC}}*R_{ij}-K_{HC}*R_{HC} & \mbox{if $R_{HC}\leq R_{ij}$} \\
			-\frac{K_{SC}*R_{SC}}{R_{SC}-R_{HC}}*R_{ij}+\frac{K_{SC}*(R_{SC})^2}{R_{SC}-R_{HC}} & \mbox{if $R_{HC}\leq R_{ij} < R_{SC}$} \\
			0 & \mbox{otherwise}. \end{array} \right .
			\label{eq:forcePP} 
\end{equation}
 where $R_{ij}$ is the distance between $i$ and $j$ cell centres,
 $K_{HC}$, $K_{SC}$ are the spring constants of hard core and elastic repulsive interactions,
 and interaction cutoffs $R_{HC}$ and $R_{SC}$ are defined as follows
 \begin{equation}
  \left \{
      \begin{array}{l}
		  	R_{SC}=R_i+R_j+(\delta_i+\delta_j) \\
		  	R_{HC}=R_i+R_j \\
    	 \end{array}
    	  \right .
    \label{eq:cutPP}
 \end{equation} 

(b) \textbf{\textit{Anterior}-\textit{anterior} cell interactions}: 

In addition to elastic and hard core repulsion forces above,
there is a spring-like attractive force acting between two anterior cells in vicinity.
 \begin{equation}
		F_{ij} = \left \{ \begin{array} {ll}
			-\frac{K_{HC}*R_{HC}-K_{SC}*R_{SC}}{R_{HC}}*R_{ij}-K_{HC}*R_{HC} & \mbox{if $ R_{HC}\leq R_{ij}$} \\
			-\frac{K_{SC}*R_{SC}}{R_{SC}-R_{HC}}*R_{ij}+\frac{K_{SC}*(R_{SC})^2}{R_{SC}-R_{HC}} & \mbox{if $R_{HC}\leq R_{ij} < R_{SC}$} \\
			\frac{K_{Attr}*R_{Attr}}{R_{Attr}-R_{Attr}}*R_{ij}-\frac{K_{Attr}*(R_{Attr})^2}{R_{Attr}-R_{SC}} & \mbox{if $R_{SC}\leq R_{ij} < R_{Attr}$} \\
			0 & \mbox{otherwise}. \end{array}
			 \right .
			\label{eq:forceAA}
\end{equation}
Here
\begin{equation}
\left \{
      \begin{array}{l}
		  	R_{Attr}=R_i+R_j+\delta_i+\delta_j \\
		  	R_{SC}=R_i+R_j+(\delta_i+\delta_j)/2 \\
		  	R_{HC}=R_i+R_j \\
    	 \end{array}
  \right .
  \label{eq:cutAA}
 \end{equation}

(c) \textbf{\textit{Anterior}-\textit{posterior} interaction:}

Finally, anterior-posterior cell pairs, similar to two posterior cells,
interact only through repulsive forces.
The interaction forces and corresponding cutoff distances for these pairs are defined through equations similar to Eq.s~(\ref{eq:forcePP})-(\ref{eq:cutPP}).

\begin{figure}[h]
  \centering \setlength{\fboxsep}{0pt}%
  \setlength{\fboxrule}{0pt}%
  \includegraphics[width=0.7\textwidth]{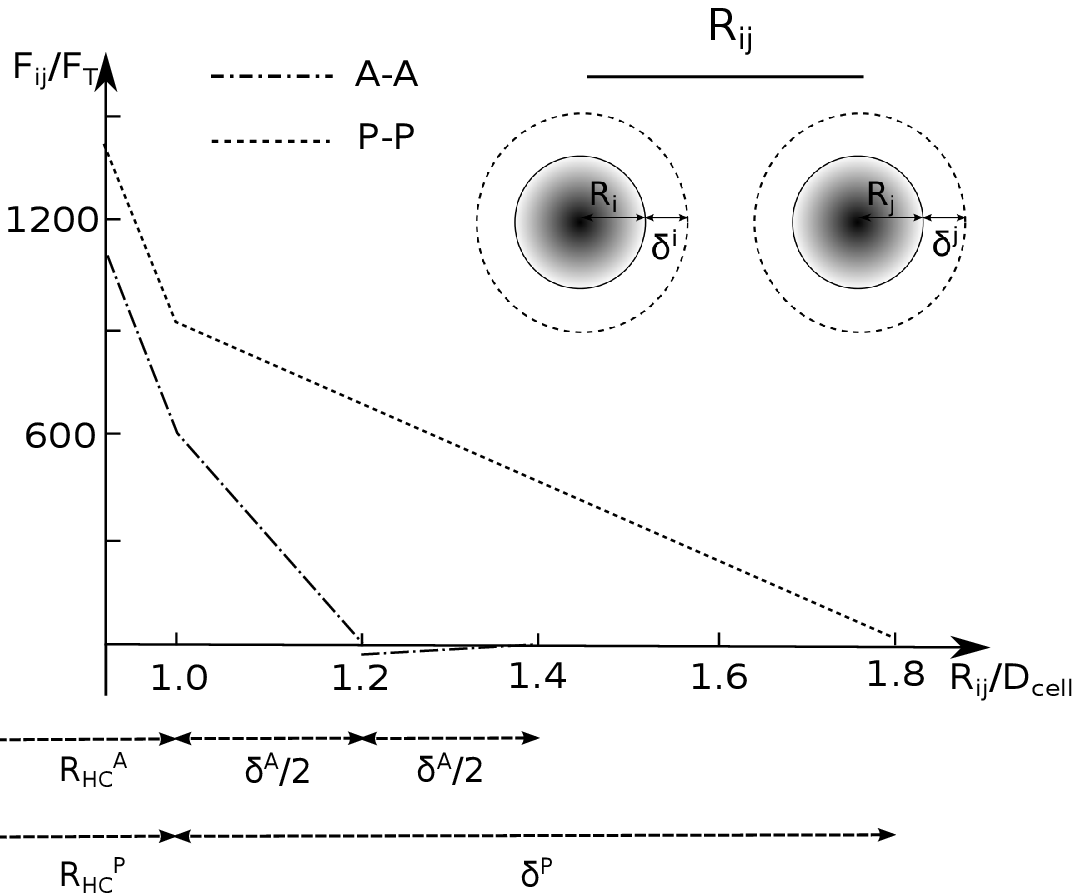}
  \caption{Interaction forces between two cells. Forces are modelled through linear springs.
  Cell-cell interaction forces depend on types of the interacting cells.
  Specifically, two anterior cells display both cell-cell adhesion and repulsion.
  Anterior-posterior or posterior-posterior cell pair interaction is entirely repulsive.
 Elastic repulsion is further subdivided into two regimes
 characterized by spring constants $K_{SC}$ and $K_{HC}$ 
to account for cell's ability to  compression to a certain degree ($K_{SC}$) as well as cell's excluded volume ($K_{HC}$).
  All numerical values are in reduced units. The unit of force is defines as $[F]=10^{-11}N$ and the length unit is the diameter of the model cells $[l]=20\mu m$
  }
  \label{fig:forces}
\end{figure}

Interaction forces between various cell pairs are shown in Fig.~\ref{fig:forces}.

\newpage

\subsection*{Parametrization}
\label{sec:params}

We choose the mass, the hard core diameter of anterior/posterior cells,
and cell's metabolic energy as the
unity of mass, length and energy correspondingly.
We assume that all cells have the same mass which stays constant throughout simulations.
Next, we set the total radii of anterior and posterior cells to $R_A^{Tot}=0.7$ and $R_P^{Tot}=0.9$ in reduced units.
Such choice of radii ensures that $\delta < R$~\cite{drasdo1995monte}
and qualitatively reproduces the gradient of PSM number density in chick embryo~\cite{benazeraf2010random} (\textit{Supplementary Information}).
The instantaneous values of cell radii change throughout simulations,
for example, during cell growth or division, however,
the ratio between R and $\delta$ is kept constant for a given cell type.
Hensen's node is considered to be a cell of the anterior type
with the hard core diameter set to 4.
The characteristic length of 
cells' density/motility gradient L which defines the length scale of the models 
is set to 25 in reduced units.

The model units can be converted to the real units once the base for the conversion in chosen.
We pick the experimental values for the mass of eukaryote cell,
the cell metabolic energy
and characteristic length of the density gradient as the conversion base.
The typical mass of an eukaryote cell is $10^{-12}kg$, which leads to the mass unit of  $[m]=10^{-12}kg$.
Following Drasdo's choice of the numerical value for the cell's metabolic energy, we define the energy unit as $[E]=F_T=10^{-16}J$~\cite{drasdo1995monte}.
Finally, the density gradient in chick embryo which has the characteristic length of about $500\mu m$~\cite{benazeraf2010random} sets the real value of the unit length $[l]$ to $20\mu m$.
Thus, model cells are roughly 2 times larger compared to the real cells.

The units of the force and the time are derived as
$[F]\sim[E]/[l]=10^{-12}N$ and  $[t]=\sqrt{\frac{[m]*[l]}{[F]}}=\sqrt{\frac{[m]*[l]^2}{[E]}}\sim\sqrt{\frac{10^{-12}*10^{-10}kg*m^2}{10^{-16}J}}=10^{-3} sec$.

Next, diffusion coefficient of posterior cells $D_P$ is set to $0.1$.
We estimate the corresponding real value to be $1.2\cdot 10^6 \mu ^2/min$,
which is roughly the $10^6$ times higher that the observed experimental value~\cite{benazeraf2010random}.
Consequently, we identify that for computational efficiency
the model dynamics is accelerated by factor of $10^6$. 
To acquire the correct time scales, one needs to substitute the
model time unit with $[t]=10^3 sec$.

Cells in anterior are less motile than in posterior.
We hypothesize that this is a consequence of an increased  cell density in the anterior.
Anterior cells form adhesive bonds with their neighbours,
creating crowded environment which, in turn, decreases their motility.
Numerically,
we assume that the anterior cells that have less than 3 neighbouring anterior cells
have the same diffusion constant as posterior cells.
Otherwise, the diffusion constant is set to $D_A=0.01$ in agreement with a factor of 10 difference between anterior and posterior cell motilities~\cite{benazeraf2010random}.
Such choice allows the anterior cells to explore their environment and decrease the motility once sufficient amount of bonds with nearest neighbours is formed.

The strengths of inter-cellular interactions are defines by spring constants $K_{HC}$, $K_{SC}$ and $K_{Attr}$
and can be derived as follows.
$K_{SC}$ is related to the Young's modulus ($Y=2*10^3 Pa$ for PSM tissues). 
The spring constant $K_{SC}$ relates to Young's modulus $Y$ as $K_{SC}\sim Y$.
Thus, $K_{SC}\sim 1000$.
$K_{HC}$ represents the hard core repulsion and should be large enough to prevent a cell to occupy other cell's excluded volume. We choose $K_{HC}=10\cdot K_{SC}$. 
$K_{Adh}$ is chosen to be about 100 times smaller that $K_{SC}$, i.e. $K_{Adh}=10$.

Finally, we parametrize the cell proliferation in the system.
Cell growth and division in our model accounts for both cell proliferation and ingression. 
Cells in caudal PSM as well as in progenitor, from where PSM cells ingress, proliferate at about rates of $10h$ order.
Thus, we assume that the the order of magnitude for the cell doubling times in our model 
is $\sim 40[t]$ (about $10h$ in real units),
leading to the cells growth rates of order $\Delta R\sim 10^{-5}$.
The actual increments in radii are drawn from a uniform distribution with the mean of $\Delta R$, to ensure asymmetry in cell cycles. 

The growth of the cells can be suppressed in crowded environments.
We assume that cell growth is suppressed
when cells are packed in the configurations
that correspond to the dense random packing configurations,
and that the cells start to grow when packing is loosens.
For threshold packing density we choose $\rho\sim 0.6$.

For computational efficiency, we limit the growth of cells to 
growth regions characterized by the lengths $GR_A$ in the anterior,
and $GR_P$ in the posterior. 
Since caudal PSM with the characteristic length $L$
has the most prominent effect on the regression of the Hensen's node,
we disregard any effects that the addition of cells in more anterior regions may have on the regression of the node and limit the growth regions in the anterior to $L$.
In the posterior, the cell proliferation is necessary to maintain the existence of cells posterior to the node throughout simulations. 
At the same time we wish to keep the overall number of cells in system relatively small for computational efficiency. 
We find that taking $GR_P=15$ was sufficient to accomplish that task. Complete list of parameters and their numerical values are presented in Table~\ref{tab:params}.

\begin{table}[!ht]
\caption{
{\bf List of simulations parameters and their values in
  dimensionless units.}}
\begin{tabular}{|l|l|l|l|l|l|l|l|}
\hline
{\bf Parameter} & {\bf Notation} & {\bf Numerical Value}\\ \hline
 Mass of PSM cell & $m$ & $1$\\ \hline
    Hard core radii of anterior/posterior cells & $R_{A}$=$R_{P}$ & 0.5\\ \hline
    Anterior cell's shell thickness & $\delta_A$ & 0.2\\ \hline
    Posterior cell's shell thickness & $\delta_P$ & 0.4\\ \hline
    Hard core radius of Hensen's node & $R_{H}$ & 2.0\\ \hline
    Cell-cell adhesion coefficient & $K_{Attr}$ & 10\\ \hline
    Cell-cell elastic repulsion coefficient & $K_{SC}$ & 1000\\ \hline
    Cell-cell hard-core repulsion coefficient  & $K_{HC}$ & 10000\\ \hline
    Anterior-posterior elastic boundary coefficient & $K_{AP}$ & 0.05\\ \hline
    Lateral elastic boundary coefficient & $K_{L}$ & 0.5 \\ \hline
    Characteristic length of graded cell density/motility & L & 25 \\ \hline
    Diffusion coefficient of anterior cells & $D_{A}$ & 0.01\\ \hline
    Diffusion coefficient of posterior cells & $D_{P}$ & 0.1\\ \hline
    Anterior growth region & $GR_{A}$ & 25\\ \hline
    Posterior growth region & $GR_{P}$ & 15\\ \hline
    Threshold packing density for cell growth & $\rho_{thresh}$ & 0.6\\ \hline
    The average increment of cells' radii & $\delta_R$ & $5\cdot 10^{-5}$\\ \hline
    Simulation time step & $\Delta t$ & 0.001\\ \hline
    Division checkpoint & $T_{div}$ & 1000\\ \hline
    Cell type update checkpoint & $T_{type}$ & $10^4$\\ \hline
\end{tabular}
 \label{tab:params}
\end{table}

\clearpage

\section*{Supplementary Figures}

\subsection*{Supplementary Figure 1}
\label{S2_Fig}

\begin{center}
 \includegraphics[width=0.7\textwidth]{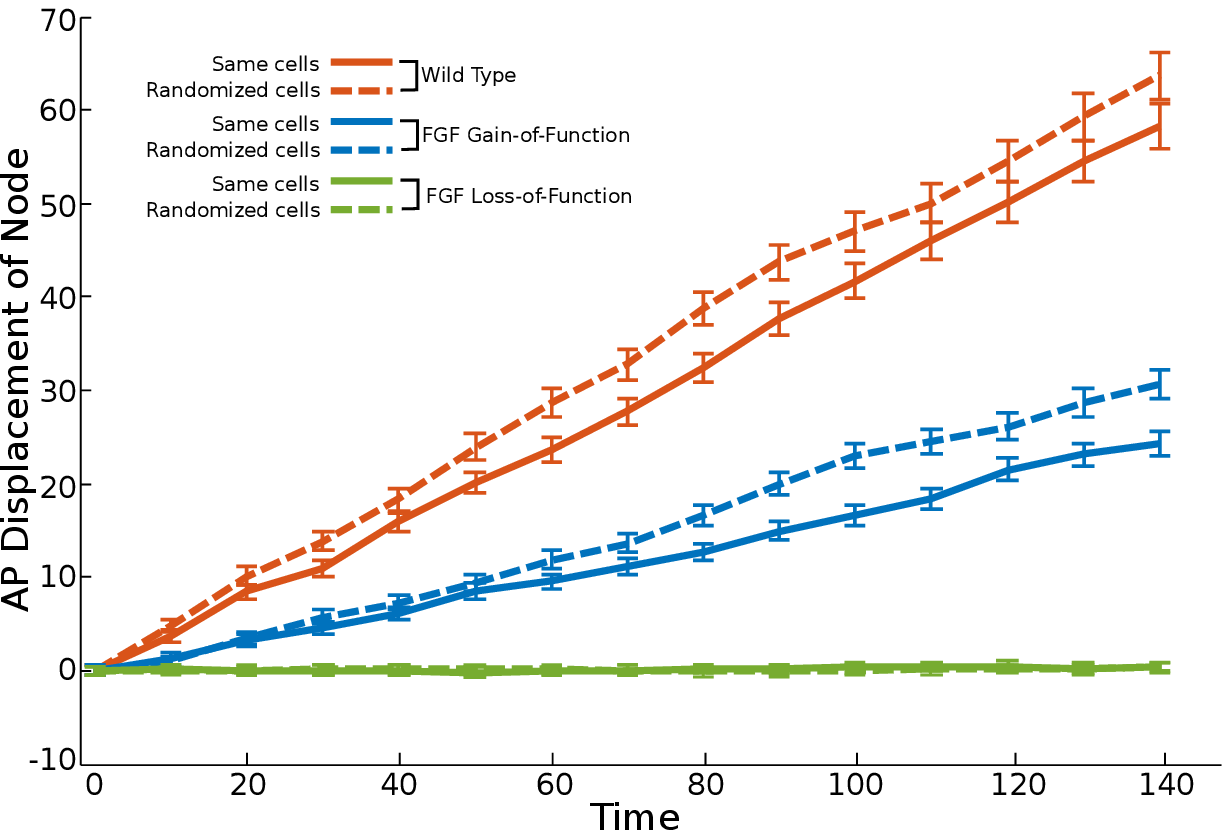}
 \end{center}
 
{\bf Randomization of cell sizes.} The effect of the randomization of cell sizes on regression of Hensen's node in wild type, FGF gain-of-function and FGF loss-of-function phenotypes. Error bars correspond to the standard deviation of node's displacement over 1000 simulation time steps.

\subsection*{Supplementary Figure 2}
\label{S3_Fig}

\begin{center}
 \includegraphics[width=0.7\textwidth]{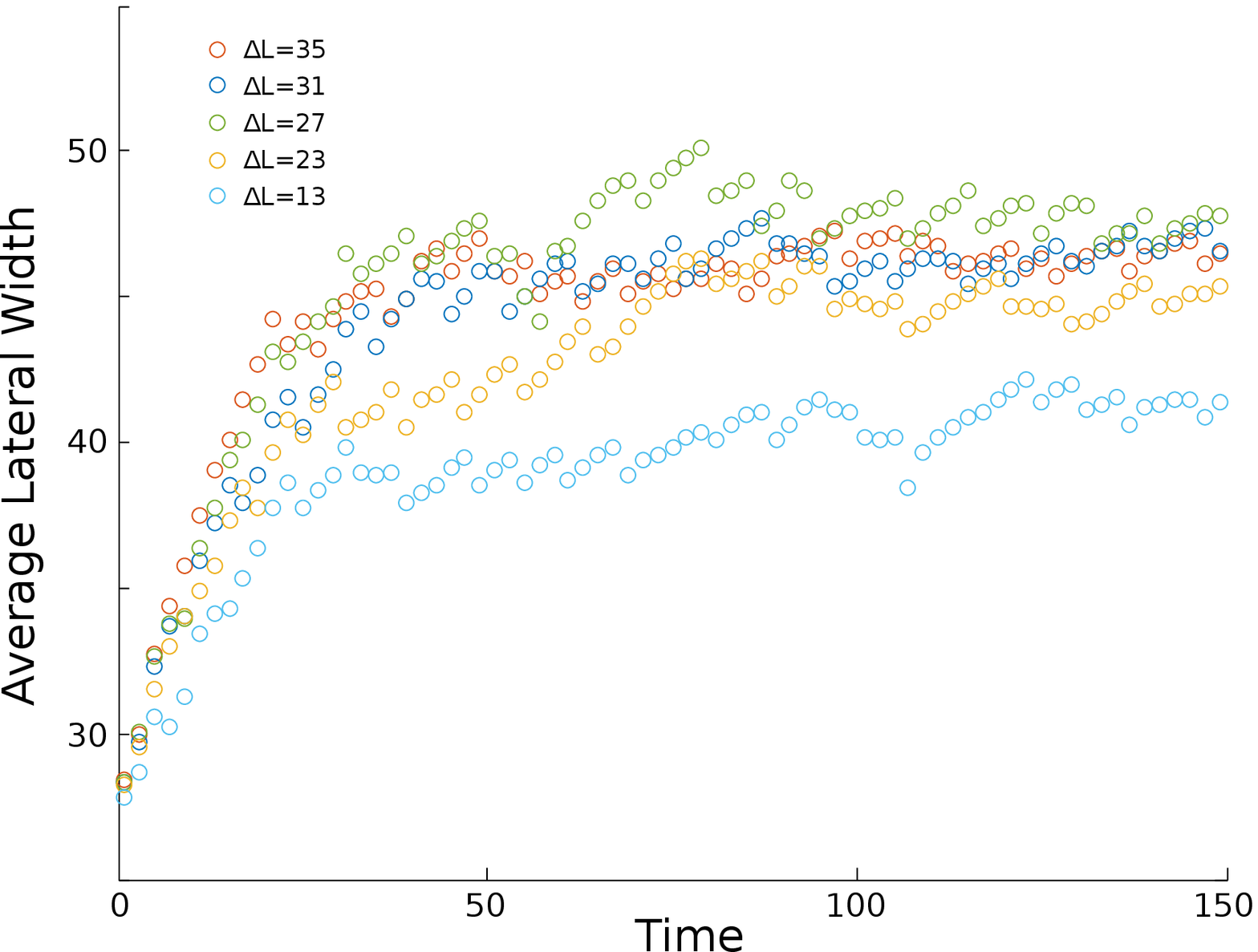}
 \end{center}
 
{\bf Effect of lateral boundary position on embryo width.} Anterior-posterior and lateral elastic spring constants are kept the same for all simulated systems ($K_{AP}$=0.005, $K_{L}$=0.1).
Simulations reveal the robustness of the model with respect to the choice of lateral boundaries.

\subsection*{Supplementary Figure 3}
\label{S4_Fig}

\begin{center}
 \includegraphics[width=0.7\textwidth]{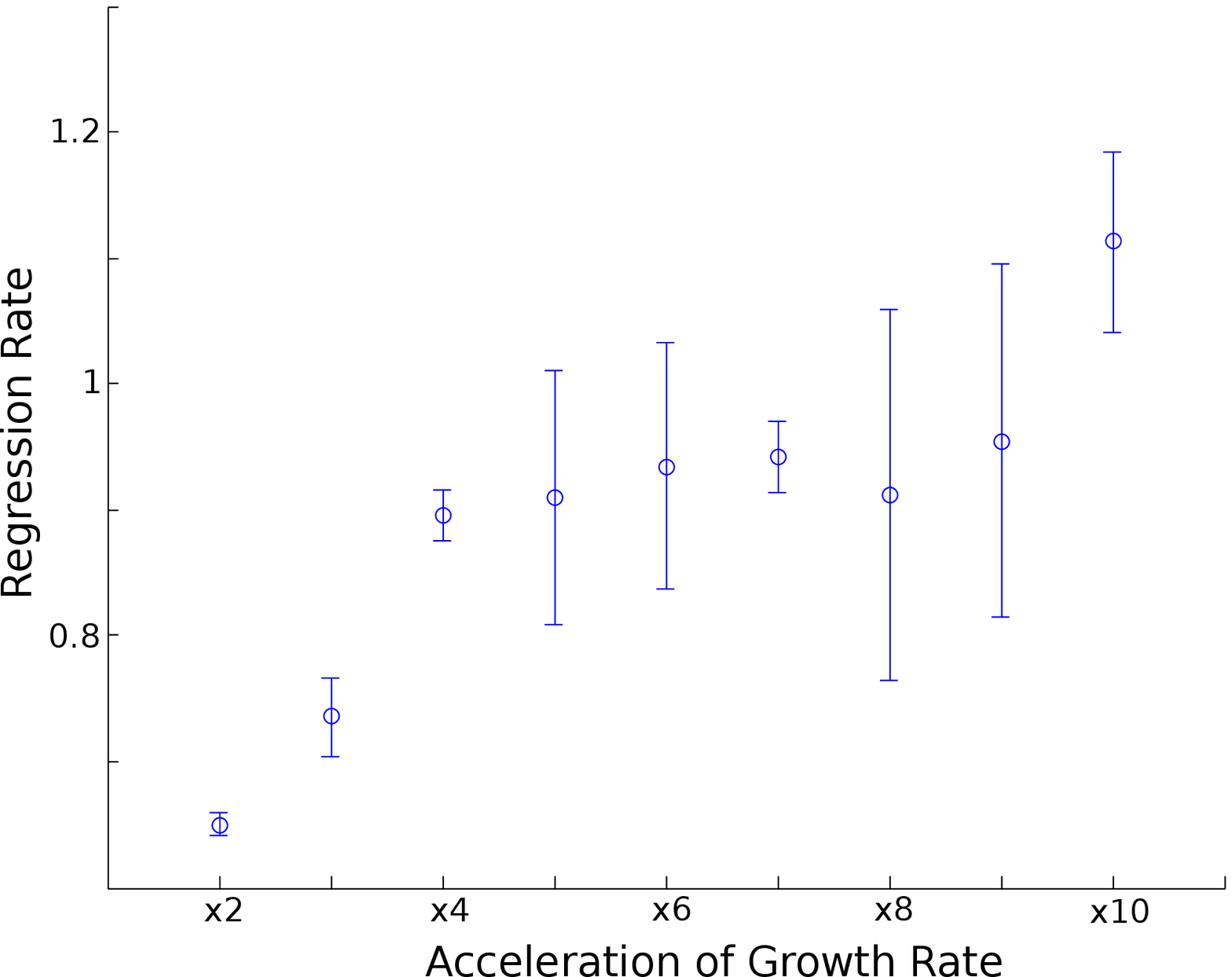}
\end{center}

{\bf Dependence of Hensen's node regression rate on the rate of cell addition.}  Regression rates are estimated for various growth rated and fitted to linear functions. Inclines of the fitted lines are plotted on Y axis. Data are averaged over two simulation runs.

\clearpage

\section*{Supplementary Movies}

Movies can be found \href{https://www.dropbox.com/sh/le514ptj97e06xj/AAD_H0oA2xu-rg11JpskrL46a?dl=0}{here}.
\subsection*{Movie 1}
\label{S1_Movie}
{\bf Simulation of the elongation of wild type embryo.} Model system consists of a mixture of anterior and posterior cells. Tissue expansion due to addition of new cells is biased towards posterior. As a result, Hensen's node and posterior cells move predominantly towards posterior.\\

\subsection*{Movie 2}
\label{S2_Movie}
{\bf Simulation of the elongation of FGF gain-of-function embryo.}  Model  system consists of entirely posterior cells. Tissue expands both anteriorly and posteriorly. The extent of node's displacement towards posterior is reduced compared to wild type embryo simulations.

\subsection*{Movie 3}
\label{S3_Movie}
{\bf Simulation of the elongation of  FGF loss-of-function model embryo.}  Model system consists of entirely anterior cells. Reduction in cell motilities prevents tissue expansion and ceases regression of the node and the elongation of embryo.

\subsection*{Movie 4}
\label{S4_Movie}
{\bf Cell number density in wild type embryo.}  Density and motility gradients are maintained over time in simulated wild type embryos. Stationary propagation of cell density and motility gradients leads to a stationary directional motion of Hensen's node towards posterior. 

\subsection*{Movie 5}
\label{S5_Movie}
{\bf Cell number density in FGF gain-of-function embryo.} Simulated FGF gain-of-function embryos are characterized by low cell number density and high cellular motility in the entire system. Flattened density and motility profiles are maintained over time and the expansion of tissue is directed both towards anterior and posterior.

\subsection*{Movie 6}
\label{S6_Movie}
{\bf Cell number density in FGF loss-of-function embryo.} Simulated FGF loss-of-function embryos are characterized by high cell number density and low cellular motility in the entire system. Flattened density and motility profiles are maintained over time. 

\subsection*{Movie 7}
\label{S7_Movie}
{\bf Simulation of the elongation of wild type embryo with randomized cell sizes.} System does not display  highly ordered packing present in the original model. The elongation of embryo and displacement of Hensen's node closely resemble system dynamics in the original model.

\subsection*{Movie 8}
\label{S8_Movie}
{\bf Simulation of the elongation of FGF gain-of-function embryo with randomized cell sizes.} Similar to wild type case, system does not display  highly ordered packing present in the original model. The elongation of embryo and displacement of Hensen's node closely follows system dynamics in the original model.

\subsection*{Movie 9}
\label{S9_Movie}
{\bf Simulation of the elongation of FGF loss-of-function embryo with randomized cell sizes.} Similar to wild type case, system does not display highly ordered packing present in the original model. The elongation of embryo and displacement of Hensen's node closely follows system dynamics in the original model.

\end {document}